\documentclass[journal]{IEEEtran}

\usepackage{amsmath,amssymb,amsfonts}
\usepackage{algorithmic}
\usepackage{array}
\usepackage[caption=false,font=normalsize,labelfont=sf,textfont=sf]{subfig}
\usepackage{textcomp}
\usepackage{stfloats}
\usepackage{url}
\usepackage{verbatim}
\usepackage{graphicx}
\def\BibTeX{{\rm B\kern-.05em{\sc i\kern-.025em b}\kern-.08em
    T\kern-.1667em\lower.7ex\hbox{E}\kern-.125emX}}
\usepackage{balance}
\usepackage{xcolor}
\usepackage{float}
\usepackage{bm}
\usepackage{cases}
\usepackage{caption}
\usepackage{subcaption}

\usepackage{algorithm}
\usepackage{algorithmic}

\usepackage{booktabs} 
\usepackage{circuitikz}
\usepackage{empheq}

\usepackage{tikz,pgf,pgfplots,pgfplotstable}
\pgfplotsset{compat=newest}
\usepgfplotslibrary{statistics}
\usepgfplotslibrary{groupplots}
\usepgfplotslibrary{fillbetween}
\usetikzlibrary{arrows.meta}

\newcommand{\win}[0]{estimation window}
\newcommand{\Win}[0]{Estimation window}
\newcommand{\Mfont}[1]{\mathrm{#1}}

\DeclareMathOperator*{\argmin}{arg\,min}

\newcommand{\Qmax}{Q_{max}}

\begin{document}

\title{Adaptive Extended Kalman Filtering for Battery State of Charge Estimation on STM32}

\author{António Barros, Edoardo Peretti, Davide Fabroni, Diego Carrera, Pasqualina Fragneto, Giacomo Boracchi}

\markboth{Barros et al. Adaptive Extended Kalman Filtering for Battery State of Charge Estimation on STM32}%
{}

\maketitle

\begin{abstract}

Accurate and computationally light algorithms for estimating the State of Charge (SoC) of a battery's cells are crucial for effective battery management on embedded systems.
In this letter, we propose an Adaptive Extended Kalman Filter (AEKF) for SoC estimation using a covariance adaptation technique based on maximum likelihood estimation - a novelty in this domain. Furthermore, we tune a key design parameter - the \win{} size - to obtain an optimal memory-performance trade-off, and experimentally demonstrate our solution achieves superior estimation accuracy with respect to existing alternative methods.
Finally, we present a fully custom implementation of the AEKF for a general-purpose low-cost STM32 microcontroller, 
showing it can be deployed with minimal computational requirements adequate for real-world usage.

\end{abstract}


\section{Introduction}

\IEEEPARstart{T}{he} ongoing energy transition is accelerating the adoption of battery-powered systems, e.g., electric vehicles, and, as such, improving the reliability of battery management systems has become a significant concern in recent years. 
Methods to track the State of Charge (SoC) of a battery cell are key components of most state of health assessment algorithms~\cite{kamali2021novel}, which are essential for ensuring effective power management, safe operation, and extending battery lifespan.
Real-time diagnostics require these algorithms to run on embedded devices, which places an extra constraint on computational power.
As a result, there has been a lot of interest in designing efficient yet computationally light algorithms that can be implemented on microcontroller units (MCUs).

A battery's cell is characterized by the total amount $\Qmax$ of charge that it can store.
We assume $\Qmax$ to be constant over time, which is reasonable across a few charge-discharge cycles.
Denoting with $Q_k$ the amount of charge stored at time $k$, the SoC $z_k$ of the cell at time $k$ is defined as
\begin{equation}
    z_k = \frac{Q_k}{\Qmax}.
\end{equation}
Directly measuring the SoC is not possible, thus each battery cell is modeled as a black-box system, which exhibits a voltage $V_k$ at its terminals according to the applied current $i_k$. 
Therefore, the goal of a SoC estimation algorithm is to estimate $z_k$, given only the measures of $i_k$ and $V_k$.

Many methods have been proposed for SoC estimation, from naïve approaches like Coulomb Counting (CC)~\cite{CC_critical} to complex ones like neural networks~\cite{lstm_soc} and electrochemical models~\cite{pybamm}. However, most of these are either inaccurate or too computationally demanding for real systems. Simple methods like CC are sensitive to wrong initialization values and neglect several hardware non-idealities, like sensor offsets, leading to a significant SoC drift. Advanced methods, such as neural networks, require extensive and often unavailable battery data, while ODE-based models are too expensive to solve in real-time on MCUs.
Thus, most viable implementations rely on an intermediate and computationally tractable solution: Kalman Filters (KFs). In this case, the battery is approximated by an Equivalent Circuit Model (ECM) with mildly non-linear dynamics, and its internal state is estimated by an Extended Kalman Filter (EKF) or its adaptive variant (AEKF).
Although more advanced filters exist, e.g., sigma-point KFs and particle filters, they have higher computational complexity and provide superior performance only for highly non-linear systems.
Since that is not the case with the aforementioned ECM, they do not offer any advantages for the considered application.

In this letter, we propose an AEKF for SoC estimation. In particular, we put forward maximum likelihood estimation (MLE) for covariance adaptation, which, although proven successful in other domains, had never been attempted in this context.
Additionally, we provide two empirical contributions: first, we analyze the optimal \win{} size, a key design criterion for this type of solution; second, we compare SoC estimation performance, both on simulated and real data, against alternative solutions for MCUs, namely CC, a base EKF and an AEKF using a covariance matching approach for covariance adaptation~\cite{cm_soc}. 
Last, we detail our implementation of the AEKF on a general-purpose STM32 MCU and analyze its performance (i.e. memory requirements, computation time, and power consumption), demonstrating its viability for real-time operation even on resource-constrained hardware.

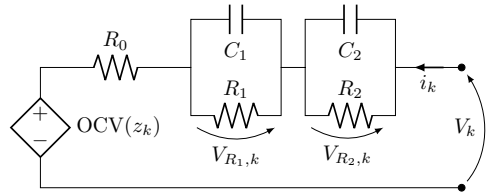
\begin{figure}[t!]
    \centering
    \scalebox{.8}{
    \begin{circuitikz}[american voltages]
\ctikzset{bipoles/resistor/height=0.3}
\ctikzset{bipoles/resistor/width=0.5}
\ctikzset{bipoles/capacitor/height=0.35}
\ctikzset{bipoles/capacitor/width=0.15}
\draw
  (1,0) to [short, *-] (-6,0)
  to [controlled voltage source, invert, l_=$\mathrm{OCV}(z_k)$, /tikz/circuitikz/bipoles/length=1.4cm] (-6,2)
  to [R, R=$R_0$] (-3.5, 2)
  to (-3.5, 2.8)
  to [C, l_=$C_1$] (-2, 2.8)
  to (-2, 2)
  to (-1.6, 2)
  to (-1.6, 2.8)
  to [C, l_=$C_2$] (-0.1, 2.8)
  to (-0.1, 2)
  to [short, -*] (1, 2)
  (-3.5, 2) to (-3.5, 1.2)
  to [R, R=$R_1$, invert] (-2, 1.2)
  to (-2,2)
  (-1.6, 2) to (-1.6, 1.2)
  to [R, R=$R_2$] (-0.1, 1.2)
  to (-0.1, 2)
  ; 

\draw [->, -latex, shorten >=0.4em, shorten <=0.4em] (1,0) to[out=50, in=-50] node[auto]{$V_k$} (1,2);
\draw [->, -latex, thick] (.7,2) -- node[auto]{$i_k$} (.2,2);
\draw [<-, latex-] (-0.3, 1) to[bend left] node[auto=left]{$V_{R_2,k}$} (-1.5, 1);
\draw [<-, latex-] (-2.1, 1) to[bend left] node[auto=left]{$V_{R_1,k}$} (-3.4, 1);
\end{circuitikz}
    }
    \caption{Improved Thévenin model of a battery's cell.}
    \vspace{-0.3cm}
    \label{fig:ecm}
\end{figure}

\section{SoC estimation algorithm}

\subsection{Second Order ECM}

We adopt a second-order ECM~\cite{aekf_ecm} for the battery model, known as the improved Thévenin model (Fig.~\ref{fig:ecm}).
The ECM employs electrical circuit elements to approximate the cell’s output voltage given a stimulus (current) and its state (stored charge).
When no load is attached, the cell exhibits a constant output voltage, designated open circuit voltage (OCV). As it heavily depends on the SoC, it is modeled as a controlled voltage source.
With a load, a current flows through the cell, and the voltage drops below the OCV. 
This drop is modeled with a resistance $R_0$ in series with the source.
Transient effects during charge or discharge, known as concentration and electrochemical polarizations, are modeled by two resistance-capacitance blocks in series: $(R_1,C_1)$ and $(R_2,C_2)$.

The parameters of the ECM are the OCV$(z_k)$ voltage source, characterized by a non-linear function of the SoC, the resistances $R_0, R_1, R_2$ and the capacitances $C_1,C_2$.
The ECM in Fig.~\ref{fig:ecm} has the following dynamics:
\begin{subequations}\label{eq:ecm_dynamics}
\begin{empheq}[left=\empheqlbrace]{align}
    z_{k+1} &= z_k + \frac{\Delta t}{\Qmax} i_k \label{eq:ecm_dynamics_z} \\
    V_{R_1,k+1} &= e^{-\frac{\Delta t}{R_1 C_1}} V_{R_1, k} + R_1 (1 - e^{- \frac{\Delta t}{R_1C_1}}) i_k  \label{eq:ecm_dynamics_v1} \\
    V_{R_2, k+1} &= e^{-\frac{\Delta t}{R_2 C_2}} V_{R_2, k} + R_2 (1 - e^{- \frac{\Delta t}{R_2 C_2}}) i_k \label{eq:ecm_dynamics_v2} \\
    V_k &= OCV(z_k) + R_0 i_k + V_{R_1, k} + V_{R_2, k} \label{eq:ecm_dynamics_v}
\end{empheq}
\end{subequations}
where $V_{R_1}$ and $V_{R_2}$ are the voltages on $R_1$ and $R_2$, $i$ and $V$ are the current and the voltage at the terminals, and $\Delta t$ is the discretization step. 
Eq.~\eqref{eq:ecm_dynamics_z} is the discrete-time Coulomb Counting equation, describing the evolution of the SoC $z_k$ by accumulating the charge supplied by the current $i$.
Eq.~\eqref{eq:ecm_dynamics_v1}-\eqref{eq:ecm_dynamics_v} 
can be derived by 
standard electric circuit analysis.

\begin{figure}[t!]
    \centering
    \includegraphics[width=\linewidth]{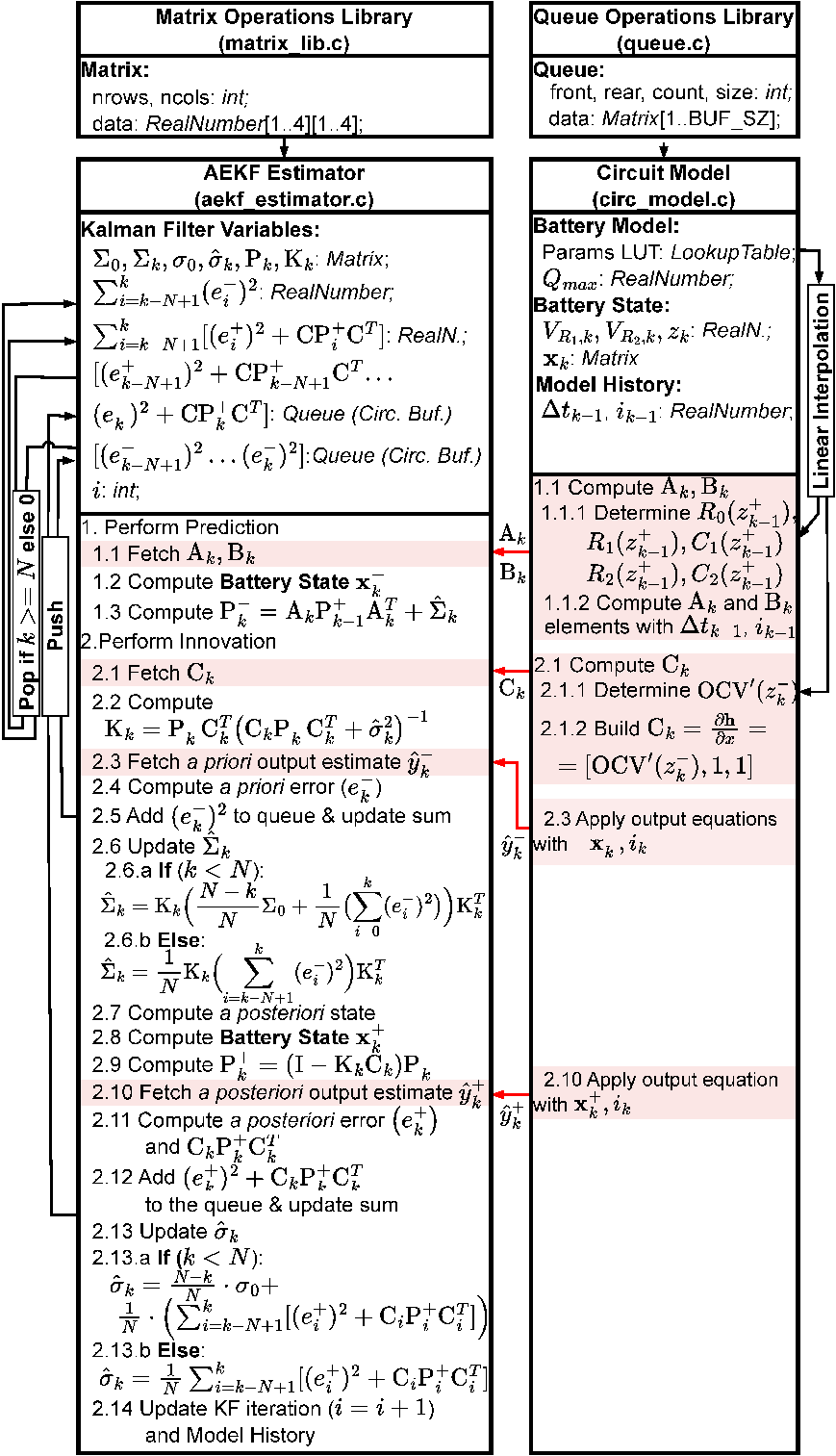}
    \vspace{-0.5cm}
    \caption{Diagram of our embedded implementation of the EKF with MLE adaptive scheme. Red arrows represents communication lines between the two modules. Black arrows represents access to modules' data structures.}
    \label{fig:diagram}
    \vspace{-0.35cm}
\end{figure}

\renewcommand{\arraystretch}{1.2} 
\begin{table}[t!]
\centering
\begin{tabular}{@{}llll@{}}
\toprule
\textbf{Window size} & \textbf{RAM} & \textbf{Time} & \textbf{Power} \\ \hline
16  & 3.55kB  & 6.16ms & 21.3mW \\
32  & 3.67kB  & 6.16ms & 21.3mW \\
64  & 3.92kB  & 6.16ms & 21.3mW \\ 
128 & 4.42kB  & 6.16ms & 21.3mW \\
\bottomrule
\end{tabular}
\caption{Resources required by our embedded implementation of the MLE scheme, for various window sizes.}
\vspace{-0.3cm}
\label{tab:stm32_req_buffer}
\end{table}

\subsection{Parameters Estimation}\label{sec:param_fitting}

The first step is to estimate the ECM parameters for a given battery cell, which consists in estimating the OCV-SoC curve and the passive components $(R_0, R_1, R_2, C_1, C_2)$. 

\subsubsection{OCV-SoC Curve Estimation}

When the current is very low, the terminal voltage $V_k$ approximates the OCV.
Therefore, we estimate the OCV-SoC relation by discharging a fully charged cell with a constant low current until the lower cut-off voltage is reached, then recharging it with a constant low current.
By continuously measuring the terminal voltage, we obtain a ``discharge OCV" and a ``charge OCV" curve.
The final OCV-SoC curve is the average of these two curves, reducing the effects of hysteresis and small voltage drops due to the battery's ohmic resistance.
In our embedded implementation, we sample the curve at equally spaced SoC points (2\% apart) to build a lookup table that is linearly interpolated at runtime. We found this setting by incrementally halving the spacing from 16\% until no performance gains could be observed.

\subsubsection{Passive Component Estimation}

Given a current profile $i_k$ and an initialization of $(R_0, R_1, R_2, C_1, C_2)$, we estimate the voltage $\hat{V}_k$, using the dynamics~\eqref{eq:ecm_dynamics} of the ECM in Fig.~\ref{fig:ecm}.
Then, all ECM parameters are estimated by solving the following non-linear least squares problem:
\begin{equation}\label{fitting_objective}
    \hat{R}_0, \hat{R}_1, \hat{R}_2, \hat{C}_1, \hat{C}_2 = 
    \argmin_{R_0,R_1,R_2,C_1,C_2} \sum_k (\hat{V}_k - V_k)^2
\end{equation}
where $V_k$ is the measured voltage, and $\hat{V}_k$ depends on $R_0,R_1,R_2,C_1,C_2$ as in \eqref{eq:ecm_dynamics}.
We use the Levenberg-Marquadt method to solve the problem in \eqref{fitting_objective}.
In particular, we adopt the \textit{incremental current} test, which consists of charging the cell for many time intervals with a constant charge current, interleaved with rest periods without any current.
This test makes the circuit's capacitive response very apparent and avoids introducing quantization errors due to discretization, given that it can be described as a sum of current steps.
In general, the passive components may depend on some external factors (e.g., temperature) or the SoC. Thus, if needed, the parameter estimation is run for a set of discrete SoC points/ temperatures, and intermediate values are interpolated at run time. In our simulations, we consider the temperature fixed and found the SoC dependence to be very weak, so we perform the above procedure only once.

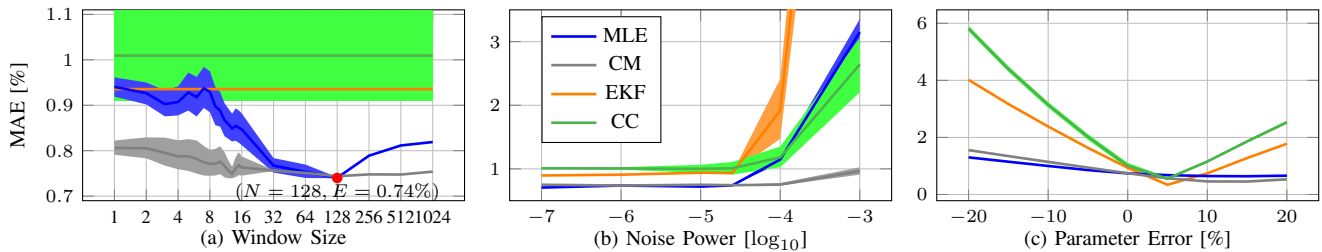
\begin{figure*}[t]
    \centering
    \begin{tikzpicture}
\def\lw{1.0pt}  
\begin{groupplot}[
    group style={group size=3 by 1, horizontal sep=.6cm}, %
    width=.28\textwidth,
    height=.14\textwidth,
    scale only axis, 
    xlabel style = {font=\footnotesize, yshift=1ex,}, 
    ylabel style = {font=\footnotesize}, 
    yticklabel style = {font = \scriptsize},
    xticklabel style = {font = \scriptsize},
]
\nextgroupplot[ 
    xlabel={(a) Window Size}, 
    ylabel={MAE [\%]}, 
    xtick = {1, 2, 4, 8, 16, 32, 64, 128, 256, 512, 1024}, 
    xticklabels = {1, 2, 4, 8, 16, 32, 64, 128, 256, 512, 1024},
    xmode = log,
    ymajorgrids,    
    xmajorgrids,
 %
    ymin=0.69,
    ymax=1.11,
]
\addplot[color=green!40!gray, line width=\lw]
    table[x=buf_size, y=cc_mean]
    {data/buf_analysis_results.txt}; 

\addplot[color=orange, line width=\lw]
    table[x=buf_size, y=ekf_mean]
    {data/buf_analysis_results.txt}; 
\addplot[color=blue, line width=\lw]
    table[x=buf_size, y=mle_mean]
    {data/buf_analysis_results.txt}; 
%
\addplot[color=gray, line width=\lw]
    table[x=buf_size, y=cm_mean]
    {data/buf_analysis_results.txt}; 
%
\draw [draw=none, fill=red, anchor=center] (axis cs:128,0.74) circle (2pt);
\node [anchor=north west, font=\scriptsize, inner sep=0pt, anchor=north] at (axis cs:128,0.73) {$(N=128, E=0.74\%)$};
%
\addplot[name path=cc_lower, draw=none,forget plot] table[x=buf_size, y=cc_lower_conf_bound]
{data/buf_analysis_results.txt}; 
\addplot[name path=cc_upper, draw=none,forget plot] table[x=buf_size, y=cc_upper_conf_bound]
{data/buf_analysis_results.txt}; 
\addplot[green!75, opacity=0.5, forget plot] fill between[of=cc_lower and cc_upper];

\addplot[name path=ekf_lower, draw=none,forget plot] table[x=buf_size, y=ekf_lower_conf_bound]
{data/buf_analysis_results.txt}; 
\addplot[name path=ekf_upper, draw=none,forget plot] table[x=buf_size, y=ekf_upper_conf_bound]
{data/buf_analysis_results.txt}; 
\addplot[orange!75, opacity=0.5, forget plot] fill between[of=ekf_lower and ekf_upper];
\addplot[name path=mle_lower, draw=none,forget plot] table[x=buf_size, y=mle_lower_conf_bound]
{data/buf_analysis_results.txt}; 
\addplot[name path=mle_upper, draw=none,forget plot] table[x=buf_size, y=mle_upper_conf_bound]
{data/buf_analysis_results.txt}; 
\addplot[blue!75, opacity=0.5, forget plot] fill between[of=mle_lower and mle_upper];
\addplot[name path=cm_lower, draw=none,forget plot] table[x=buf_size, y=cm_lower_conf_bound]
{data/buf_analysis_results.txt}; 
\addplot[name path=cm_upper, draw=none,forget plot] table[x=buf_size, y=cm_upper_conf_bound]
{data/buf_analysis_results.txt}; 
\addplot[gray!75, opacity=0.5, forget plot] fill between[of=cm_lower and cm_upper];
\nextgroupplot[
    xlabel={(b) Noise Power [$\log_{10}$]}, 
    ymajorgrids,    
    xmajorgrids,
 %
    ymin=0.5,
    ymax=3.5,
    legend style={
            at={(axis cs: -7,3.4)},   
            anchor=north west,          
    	font=\footnotesize,                
    },
]
\addplot[color=blue, line width=\lw]
    table[x=log10_var, y=mle_mean]
    {data/noise_analysis_results.txt}; 
    \addlegendentry{MLE}
\addplot[color=gray, line width=\lw]
    table[x=log10_var, y=cm_mean]
    {data/noise_analysis_results.txt}; 
    \addlegendentry{CM}
\addplot[color=orange, line width=\lw]
    table[x=log10_var, y=ekf_mean]
    {data/noise_analysis_results.txt}; 
    \addlegendentry{EKF}
\addplot[color=green!40!gray, line width=\lw]
    table[x=log10_var, y=cc_mean]
    {data/noise_analysis_results.txt}; 
    \addlegendentry{CC}
%
\addplot[name path=ekf_lower, draw=none,forget plot] table[x=log10_var, y=ekf_lower_conf_bound]
{data/noise_analysis_results.txt}; 
\addplot[name path=ekf_upper, draw=none,forget plot] table[x=log10_var, y=ekf_upper_conf_bound]
{data/noise_analysis_results.txt}; 
\addplot[orange!75, opacity=0.5, forget plot] fill between[of=ekf_lower and ekf_upper];
\addplot[name path=mle_lower, draw=none,forget plot] table[x=log10_var, y=mle_lower_conf_bound]
{data/noise_analysis_results.txt}; 
\addplot[name path=mle_upper, draw=none,forget plot] table[x=log10_var, y=mle_upper_conf_bound]
{data/noise_analysis_results.txt}; 
\addplot[blue!75, opacity=0.5, forget plot] fill between[of=mle_lower and mle_upper];
\addplot[name path=cm_lower, draw=none,forget plot] table[x=log10_var, y=cm_lower_conf_bound]
{data/noise_analysis_results.txt}; 
\addplot[name path=cm_upper, draw=none,forget plot] table[x=log10_var, y=cm_upper_conf_bound]
{data/noise_analysis_results.txt}; 
\addplot[gray!75, opacity=0.5, forget plot] fill between[of=cm_lower and cm_upper];
\addplot[name path=cc_lower, draw=none,forget plot] table[x=log10_var, y=cc_lower_conf_bound]
{data/noise_analysis_results.txt}; 
\addplot[name path=cc_upper, draw=none,forget plot] table[x=log10_var, y=cc_upper_conf_bound]
{data/noise_analysis_results.txt}; 
\addplot[green!75, opacity=0.5, forget plot] fill between[of=cc_lower and cc_upper];
\nextgroupplot[
    xlabel={(c) Parameter Error [\%]}, 
    ymajorgrids,    
    xmajorgrids,
]
\addplot[color=orange, line width=\lw]
    table[x=err_per, y=ekf_mean]
    {data/param_err_analysis_results.txt}; 
%
\addplot[color=blue, line width=\lw]
    table[x=err_per, y=mle_mean]
    {data/param_err_analysis_results.txt}; 
%
\addplot[color=gray, line width=\lw]
    table[x=err_per, y=cm_mean]
    {data/param_err_analysis_results.txt}; 
%
\addplot[color=green!40!gray, line width=\lw]
    table[x=err_per, y=cc_mean]
    {data/param_err_analysis_results.txt}; 
%
\addplot[name path=ekf_lower, draw=none,forget plot] table[x=err_per, y=ekf_lower_conf_bound]
{data/param_err_analysis_results.txt}; 
\addplot[name path=ekf_upper, draw=none,forget plot] table[x=err_per, y=ekf_upper_conf_bound]
{data/param_err_analysis_results.txt}; 
\addplot[orange!75, opacity=0.5, forget plot] fill between[of=ekf_lower and ekf_upper];
\addplot[name path=mle_lower, draw=none,forget plot] table[x=err_per, y=mle_lower_conf_bound]
{data/param_err_analysis_results.txt}; 
\addplot[name path=mle_upper, draw=none,forget plot] table[x=err_per, y=mle_upper_conf_bound]
{data/param_err_analysis_results.txt}; 
\addplot[blue!75, opacity=0.5, forget plot] fill between[of=mle_lower and mle_upper];
\addplot[name path=cm_lower, draw=none,forget plot] table[x=err_per, y=cm_lower_conf_bound]
{data/param_err_analysis_results.txt}; 
\addplot[name path=cm_upper, draw=none,forget plot] table[x=err_per, y=cm_upper_conf_bound]
{data/param_err_analysis_results.txt}; 
\addplot[gray!75, opacity=0.5, forget plot] fill between[of=cm_lower and cm_upper];
\addplot[name path=cc_lower, draw=none,forget plot] table[x=err_per, y=cc_lower_conf_bound]
{data/param_err_analysis_results.txt}; 
\addplot[name path=cc_upper, draw=none,forget plot] table[x=err_per, y=cc_upper_conf_bound]
{data/param_err_analysis_results.txt}; 
\addplot[green!75, opacity=0.5, forget plot] fill between[of=cc_lower and cc_upper];
\end{groupplot}
\end{tikzpicture}%
    \vspace{-0.2cm}
    \caption{95\% MAE confidence bands for synthetic data against (a) window size, (b) noise power and (c) parameters error.}
    \vspace{-0.2cm}
    \label{fig:grouplot}
\end{figure*}

\subsection{AEKF Based on the Maximum Likelihood Principle}

Our variable of interest, the SoC $z_k$, is one of the states of the dynamical system in~\eqref{eq:ecm_dynamics}.
The problem of estimating the states of a dynamical system is known as filtering.
A general discrete-time dynamical system, with linear state transition, non-linear output equation, and Gaussian noise, is given by:
\begin{subequations}\label{eq:sys_dynamics}
\begin{empheq}[left=\empheqlbrace]{align}
    \mathbf{x}_k &= \Mfont{A} \mathbf{x}_{k-1} + \Mfont{B} u_{k-1} + w_k \ &  \hfill w_k \sim \mathcal{N}(0,\Sigma) \label{eq:sys_dynamics_state} \\
    y_k &= h(\mathbf{x}_k) + \Mfont{D}u_k + v_k & v_k \sim \mathcal{N}(0, \sigma^2) \label{eq:sys_dynamics_out}
\end{empheq}
\end{subequations}
In our case, the input is $u = i$, the output is $y = V$ and the state is $\mathbf{x} = (z, V_{R_1}, V_{R_2})^T$.
Comparing the two systems in~\eqref{eq:ecm_dynamics} and~\eqref{eq:sys_dynamics}, one can express $\Mfont{A,B,D},h$ in terms of the ECM parameters.
The KF is the optimal state estimator in case of linear systems. However, since the output function $h$ is non-linear (being OCV non-linear), we require an Extended KF,
which replaces $h$ with its first order approximation given by the matrix $\Mfont{C} = \frac{\partial h}{\partial \mathbf{x}} = \begin{bmatrix}
        OCV'(z) & 1 & 1
    \end{bmatrix}$.
    
The noise covariances, which heavily affect the filter's performance, are notoriously hard to set.
Heuristics-based selection criteria are often sub-optimal, even after extensive manual fine-tuning. 
Besides, the covariances can change over time if the system's parameters fluctuate, for example, with varying operating conditions or aging. 
The adaptive EKFs mitigate this issue by adjusting the covariances at runtime. 
The most common approach is based on covariance matching (CM) and has already been applied to SoC estimation~\cite{cm_soc}. 
More recently, a superior covariance adaptation strategy based on MLE has been proposed for a completely different task - spacecraft navigation~\cite{mle_aekf_space}. 
In this letter, we investigate this scheme for SoC estimation. 
As proved in~\cite{mle_aekf_space}, the covariance matrices can be iteratively updated based on a window of past filter iterations. Using a sliding window of the previous \(N\) samples, in our case the update equations become:
\begin{equation}
\label{eq:mle_cov_update}
\begin{aligned}
    \hat{\Sigma}_k &= \Mfont{K}_k \frac{1}{N} \sum_{i=k-N+1}^k \left[ (e_i^{-})^2 \right] \Mfont{K}_k^T
    \\
    \hat{\sigma}_k &= \frac{1}{N} \sum_{i=k-N+1}^k \left[ (e_i^{+})^2 + \Mfont{C} \Mfont{P}_i^{+} \Mfont{C}^T \right]
\end{aligned}
\end{equation}
where \(e_i^-\) and \(e_i^+\) are the residuals after prediction and correction, $\Mfont{K}_i$ is the Kalman gain, and \(\Mfont{P}_i^{+}\) is the state estimate covariance, all at \mbox{time-step \(i\)}.
More details on these quantities can be found in Fig.~\ref{fig:diagram}.

As made evident in (\ref{eq:mle_cov_update}), the adaptive algorithm depends on the size $N$ of the sliding window. 
A large $N$  potentially offers better performance but is more costly memory-wise. 
Given the extremely limited memory of the target platform, the size of $N$ constitutes one of the main bottlenecks of our implementation.
In Section~\ref{sec:experiments}, we extend the analysis in~\cite{mle_aekf_space}, and we look into how this parameter affects performance and how a size was chosen to achieve an optimal trade-off.

\subsection{Embedded Implementation and Computational Complexity}

We devised a fully custom C-language implementation of the MLE scheme and tested it on an STM32 NUCLEO-G071RB development board.
A diagram of the implementation is shown in Fig.~\ref{fig:diagram}. Each major block represents a different C module, responsible for aggregating both data structures — listed immediately below each module's name — and functions. 
The two main ones are the \texttt{circ\_model.c}, responsible for simulating the circuit dynamics, and the \texttt{aekf\_estimator.c}, implementing the filter. 
Below the data structures is also reported a detailed sequence of steps corresponding to the algorithm's core. 
Red arrows signal communications between the modules, while black arrows denote data flow within the module involving access to one of its data structures.
Two other modules encapsulate the main auxiliary libraries: \texttt{matrix\_lib.c} implements core algebraic matrix operations, \texttt{queue\_lib.c} implements a queue data structure used to store past filter covariance matrices. 

Table~\ref{tab:stm32_req_buffer} shows the memory requirements, the computation time, and the power consumption (averaged over a period of 10s) for one filter iteration as a function of window size with the internal clock speed of the MCU set to 64MHz.
RAM values are the ones reported by STM32CubeIDE after compilation of the binary.
In comparison, the EKF requires 3.2kB, while CM has the same memory requirement as MLE.
The computation time was measured using an internal, highly precise hardware timer of the MCU, which was (re)initialized before each filter iteration and reset immediately afterward.
Power was measured with an external X-NUCLEO-LPM01A programmable power supply - configured to output 3.3V - and its STM32CubeMonPwr
software companion. 
As shown in Fig.~\ref{fig:diagram}, we adopt circular buffers to store previous filter's states, which allows us to update the covariances in~\eqref{eq:mle_cov_update} in constant time. Thus, execution time and power consumption do not depend on the window size.
We observe that deploying our implementation on the entry-level STM32 MCU with 32kB of RAM and a window size of 128 samples enables us to manage up to 7 cells with a single unit. Furthermore, considering a typical sampling interval of hundreds of milliseconds for the current and voltage measurements, its per-iteration execution time is well within real-time requirements.

\begin{figure}[t]
    \centering 
    \pgfplotsset{scaled x ticks=false}
\begin{tikzpicture}
\def\lw{1.0pt}  
\begin{groupplot}[
    group style={group size=2 by 1, horizontal sep=.6cm}, %
    width=.42\columnwidth,
    height=.33\columnwidth,
    title style = {font=\small, at={(0.5, 0.9)}, },
    scale only axis, %
    xlabel style = {font=\footnotesize, yshift=1ex,}, 
    ylabel style = {font=\footnotesize}, 
    yticklabel style = {font = \scriptsize},
    xticklabel style = {font = \scriptsize},
]
\nextgroupplot[ 
    title={Discharging}, 
    xlabel={Time [min]}, 
    ylabel={SoC [\%]}, 
    ylabel style = {at={(-0.1, 0.5)}},
    xtick = {0, 1800, 3600, 5400, 7200, 9000, 10800, 12600, 14400}, 
    xticklabels = {0, 30, 60, 90, 120, 150, 180, 210, 240},
    ymajorgrids,    
    xmajorgrids,
    xmin=0,
    xmax=9500,
    ymin=20,
    ymax=105,
]
\addplot[color=green!40!gray, line width=\lw]
    table[each nth point={10}, x=t, y=mle]
    {data/discharge_3.5A_results_fast.txt}; 
\addplot[color=gray, line width=\lw]
    table[each nth point={10}, x=t, y=cm]
    {data/discharge_3.5A_results_fast.txt}; 
\addplot[color=orange, line width=\lw]
    table[each nth point={10}, x=t, y=ekf]
    {data/discharge_3.5A_results_fast.txt}; 
\addplot[color=blue, line width=\lw]
    table[each nth point={10}, x=t, y=cc]
    {data/discharge_3.5A_results_fast.txt}; 
\addplot[color=black, dashed, line width=\lw]
    table[unbounded coords=jump, each nth point={10}, x=t, y=soc_gt]
    {data/discharge_3.5A_results_fast.txt}; 
\nextgroupplot[ 
    title={Charging}, 
    xlabel={Time [min]}, 
    xtick = {0, 1800, 3600, 5400, 7200, 9000, 10800, 12600, 14400}, 
    xticklabels = {0, 30, 60, 90, 120, 150, 180, 210, 240},
    ymajorgrids,    
    xmajorgrids,
    xmin=0,
    xmax=9580,
    ymin=20,
    ymax=48,
    legend columns=2,
    legend style={
        at={(axis cs: 9500,20)},   
        anchor=south east,
    	font=\scriptsize,                
    },
]
\addplot[color=blue, line width=\lw]
    table[each nth point={10}, x=t, y=cc]
    {data/charge_0.9A_results_fast.txt}; 
    \addlegendentry{CC}

\addplot[color=orange, line width=\lw]
    table[each nth point={10}, x=t, y=ekf]
    {data/charge_0.9A_results_fast.txt}; 
    \addlegendentry{EKF}

\addplot[color=gray, line width=\lw]
    table[each nth point={10}, x=t, y=cm]
    {data/charge_0.9A_results_fast.txt}; 
    \addlegendentry{CM}

\addplot[color=green!40!gray, line width=\lw]
    table[each nth point={10}, x=t, y=mle]
    {data/charge_0.9A_results_fast.txt}; 
    \addlegendentry{MLE}

\addplot[color=black, dashed, line width=\lw]
    table[unbounded coords=jump, each nth point={10}, x=t, y=soc_gt]
    {data/charge_0.9A_results_fast.txt}; 
    \addlegendentry{GT}
\end{groupplot}
\end{tikzpicture}%
    \caption{SoC estimation results on real data.}
    \vspace{-0.2cm}
    \label{fig:real_data}
\end{figure}
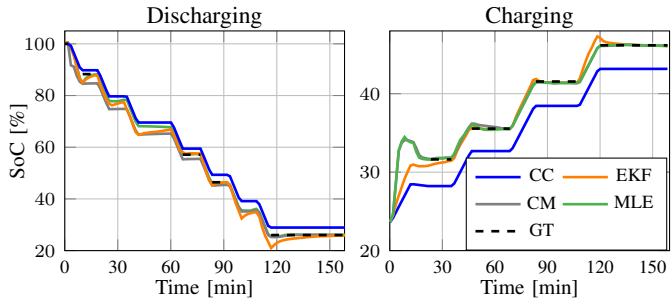

\section{Experiments}
\label{sec:experiments}

In this section, we evaluate our MLE AEKF, using both simulated and real data acquisitions.
We compare the MLE and CM adaptive EKFs with the naïve Coulomb Counting (CC) and the plain EKF.
We measure estimation performance as the point-wise absolute error between the estimated SoC and the ground truth, which in turn yields the mean absolute error (MAE) following temporal averaging.

\subsection{Experiments on Simulated Data}

We generate synthetic data using PyBaMM~\cite{pybamm}, a state-of-the-art Python library for battery simulation, offering various electrochemical models and their parametrizations, along with general differential equation solvers. For our experiments, we adopt the Doyle-Fuller-Newman battery model with parameters emulating an LG M50 cell. We generated synthetic data using the WLTC profile, which simulates complex charge/discharge phases to stress an electric vehicle's battery in various driving conditions. 
As described in Section~\ref{sec:param_fitting}, we fit \textit{SoC-independent} ECM parameters, since we observe that this model is accurate enough to match simulated data, with ground truth provided by the simulator.
Furthermore, we report the $95\%$ confidence interval computed from 1000 Monte Carlo simulations using different noise realizations.

\subsubsection{\Win{} size}
Both the MLE and the CM schemes adapt the filtering using a sliding window of past filter samples.
Since memory represents a strong limitation for MCUs, we first analyze the window size.
Fig.~\ref{fig:grouplot} shows, for all methods, the MAE as a function of this parameter.
We can observe that both the MLE and the CM schemes achieve their minimum with a window size of $N=128$.
For all subsequent experiments, we adopt this size.

\subsubsection{Robustness to noise power}
We assess the robustness of the filtering algorithms with respect to AWGN affecting the measured input current and output voltage.
Fig.~\ref{fig:grouplot}b shows the MAE for increasing noise power.
We can observe that, for reasonable noise powers, the MLE and CM schemes are equivalent, and both outperform the baselines.

\subsubsection{Robustness to modeling errors or to parameters that are changing over time}
Imprecise fitting might result in errors in the estimated ECM parameters.
Fig.~\ref{fig:grouplot}c shows the MAE as a function of the relative errors in the ECM parameters.
We can observe that the MLE and the CM are more robust for both positive and negative errors.

\subsection{Experiments on Real Data}

We evaluate our algorithm on real data acquired with a STEVAL-L99615C board from five LG INR18650MJ1 cells.
In this case, we fit \textit{SoC-dependent} ECM parameters.
For evaluation, we consider one discharging and two charging profiles.
The OCV is approximately equal to the measured terminal voltage during rest periods, since the current is zero.
Therefore, we take the ground truth as the inverse of the OCV function at the terminal voltage $V_k$, and we evaluate the performance only during rest periods.

Fig.~\ref{fig:real_data} shows the results for one cell on two profiles and Table~\ref{tab:real_data} reports the MAE of the three profiles, averaged over the five cells.
The CC suffers from a severe offset, while the EKF can match the ground truth but has strong overshooting.
The MLE scheme can precisely match the ground truth, with greater precision than the CM, effectively achieving 
comparable or significantly lower estimation error.

\begin{table}[t]
\centering
\begin{tabular}{@{}llll@{}}
\toprule
\textbf{Profile} & \textbf{EKF} & \textbf{MLE} & \textbf{CM} \\ \hline
Charging 1               & 0.472\%       & 0.175\%      & 0.172\%       \\
Charging 2               & 0.410\%       & 0.160\%      & 0.240\%       \\
Discharging 1               & 0.785\%       & 0.190\%      & 0.485\%       \\
\bottomrule
\end{tabular}
\caption{MAE averaged over five cells for the three profiles.} 
\vspace{-0.2cm}
\label{tab:real_data}
\end{table}

\section{Conclusion}

We investigated SoC estimation algorithms suitable for implementation on embedded devices. Given the target platform, we focused on EKF-based solutions, comparing the CM and MLE adaptive schemes, introducing the latter as a novelty in the battery management literature. Although they perform equivalently on simulated data, the MLE scheme proved superior on real data, for which its optimal covariance adaption better accounts for more significant model inaccuracies. With a fully custom embedded implementation, we also demonstrated it is suitable for real-world deployment on STM32 MCUs. In future work, we plan to adopt the MLE scheme in a combined SoC and state of health estimation \cite{kamali2021novel}.

\bibliographystyle{IEEEtran}
\bibliography{ref.bib}

\end{document}